\documentclass[journal=jacsat,manuscript=article]{achemso}

\usepackage{graphicx}
\usepackage{caption}
\usepackage{subcaption}
\usepackage{comment}
\usepackage{hyperref}
\usepackage{gensymb}

\usepackage{subfiles}
\usepackage{mathtools}  
\usepackage{amssymb}    
\usepackage{braket}
\usepackage{amsthm}
\usepackage{siunitx}
\usepackage{xcolor}
\DeclareUnicodeCharacter{2212}{\ensuremath{-}}
\usepackage{stackengine}
\newcommand{\op}[1]{\mathop{}\!\ensurestackMath{\stackon[-.95ex]{%
  {#1}}{\smash{{\hat{}}}}}}
\newcommand{\opdag}[1]{\mathop{}\!\op{#1}^{\dag}}

\hypersetup{colorlinks = true, citebordercolor={blue}, linkcolor={blue}, citecolor={blue}, urlcolor={blue}}

\setcitestyle{super}
\usepackage{xr}

\author{Lillian I. Payne Torres}
\affiliation{Department of Chemistry and The James Franck Institute, The University of Chicago, Chicago, IL 60637}
\author{Anna O. Schouten}
\affiliation{Department of Chemistry and The James Franck Institute, The University of Chicago, Chicago, IL 60637}
\author{LeeAnn M. Sager-Smith}
\affiliation{Department of Chemistry, Saint Mary's College, Notre Dame, IN 46556}
\author{David A. Mazziotti}
\email{damazz@uchicago.edu}
\affiliation{Department of Chemistry and The James Franck Institute, The University of Chicago, Chicago, IL 60637}

\title{A Molecular Perspective of Exciton Condensation from Particle-hole Reduced Density Matrices}

\begin{document}



\begin{tocentry}
\centering
\includegraphics[height=4.5cm]{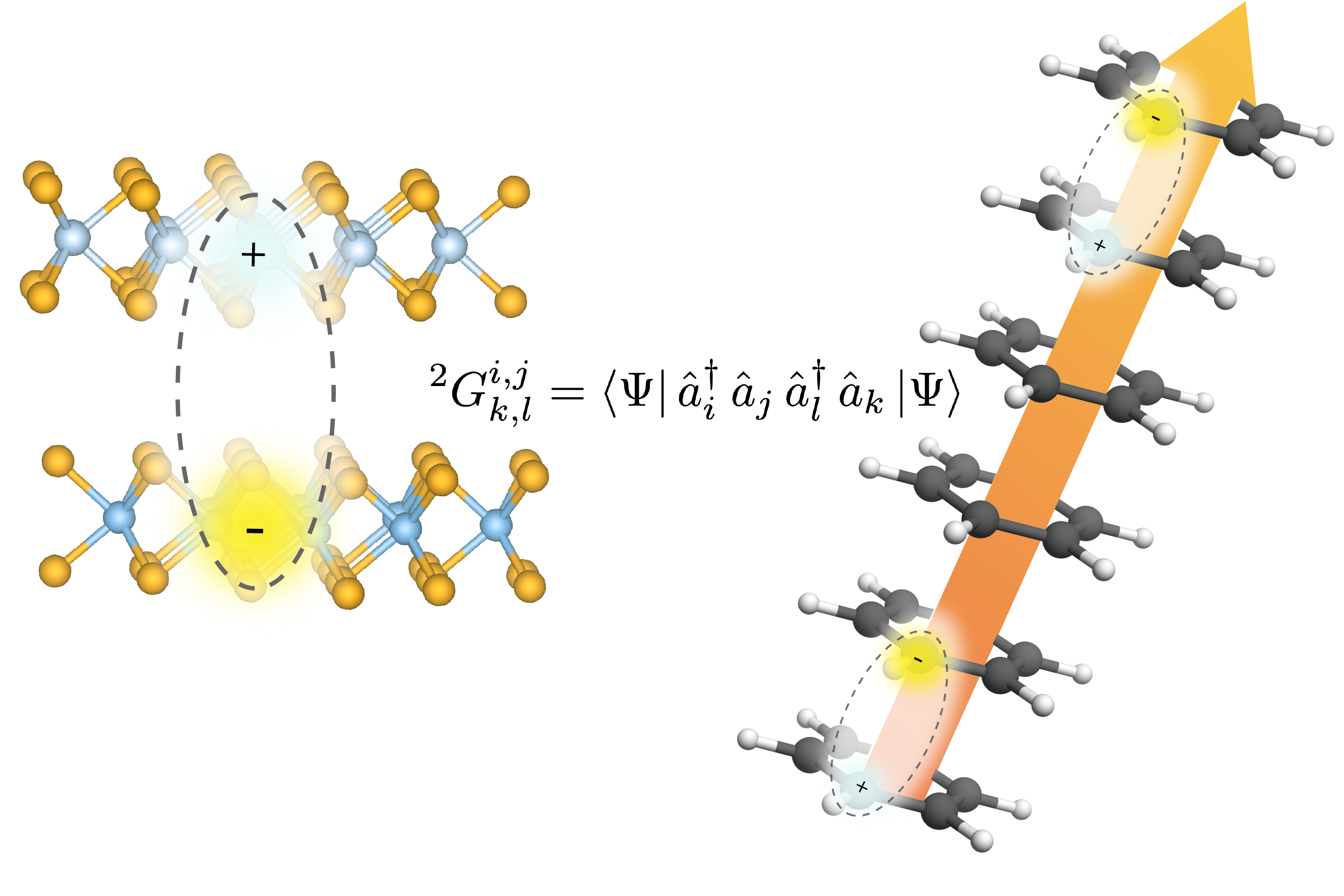}
\end{tocentry}
\begin{abstract}
Exciton condensation, the Bose-Einstein-like condensation of quasibosonic particle-hole pairs, has been the subject of much theoretical and experimental interest and holds promise for ultra-energy-efficient technologies. Recent advances in bilayer systems, such as transition metal dichalcogenide heterostructures, have brought us closer to the experimental realization of exciton condensation without the need for high magnetic fields. In this perspective, we explore progress towards understanding and realizing exciton condensation, with a particular focus on the characteristic theoretical signature of exciton condensation: an eigenvalue greater than one in the particle-hole reduced density matrix, which signifies off-diagonal long-range order. This metric bridges the gap between theoretical predictions and experimental realizations by providing a unifying framework that connects exciton condensation to related phenomena, such as Bose-Einstein condensation and superconductivity. Furthermore, our molecular approach integrates exciton condensation with broader excitonic phenomena, including exciton-related entanglement and correlation, unlocking potential advancements in fields like quantum materials and energy transport. We discuss connections between recent experimental and theoretical work and highlight the discoveries that may arise from approaching exciton condensation from a molecular perspective.
\end{abstract}


\section{Introduction}
Excitons, quasibosonic bound states consisting of an electron-hole pair, are instrumental to many important molecular and material properties and phenomena. While excitons do not transfer mass or charge, mobile excitons can transport energy via their excitation energy, a process that is fundamental to highly efficient energy transfer in photosynthesis\cite{schouten_exciton-condensate-like_2023,adolphs_how_2006} and leveraged for the design of optoelectronic devices.\cite{dijkstra_efficient_2019,mattioni_design_2021,giannini_exciton_2022,mueller_exciton_2018,high_spontaneous_2012} As excitons are quasibosonic, they can undergo a Bose-Einstein condensation into a single quantum state, which may be capable of dissipationless energy transfer.\cite{fil_electron-hole_2018,keldysh_coherent_2017}

Exciton condensation is related to traditional superconductivity, which is the condensation of pairs of fermions. While researchers have long struggled to realize superconductivity at temperatures barely approaching room temperature,\cite{zhou_high-temperature_2021} exciton condensation is predicted to be possible at higher temperatures than traditional superconductivity because excitons are both less massive and more-tightly bound than Cooper pairs.\cite{fuhrer_chasing_2016}
This potential for dissipationless energy transfer at higher temperatures has made the realization of exciton condensation a very desirable goal.\cite{eisenstein_boseeinstein_2004,snoke_boseeinstein_2014,butov_macroscopically_2002,butov_exciton_2003,kasprzak_boseeinstein_2006,min_room-temperature_2008,kharitonov_electron_2008, high_condensation_2012,high_spontaneous_2012,li_excitonic_2017,liu_quantum_2017,wang_evidence_2019,sigl_signatures_2020,ma_strongly_2021,gao_evidence_2023} However, due to fleeting lifetimes of excitons at room temperature, experimental realization of exciton condensation has been challenging.\cite{kalt_excitonic_2024} Nonetheless, scientists have made progress in preparing exciton condensation through the coupling of excitons to polaritons\cite{kasprzak_boseeinstein_2006} and via spatial separation of electrons and holes in bilayer materials such as semiconductor quantum wells and bilayers of van der Waals heterostructures.\cite{eisenstein_boseeinstein_2004,high_condensation_2012,min_room-temperature_2008,kogar_signatures_2017,li_excitonic_2017,wang_evidence_2019}

 The emergence of any quantum condensation in a system indicates the presence of long-range correlations characterized by off-diagonal long-range order (ODLRO) in the appropriate reduced density matrix (RDM).\cite{yang_concept_1962,penrose_bose-einstein_1956,garrod_particle-hole_1969,safaei_quantum_2018} In the case of exciton condensation, ODLRO is characterized by the presence of off-diagonal elements in the particle-hole RDM.\cite{safaei_quantum_2018,garrod_particle-hole_1969} Off-diagonal long-range order in exciton condensates results in an eigenvalue in the particle-hole reduced density matrix that is greater than one, indicating the presence of more than one particle-hole pair in a given excitonic mode and hence, the presence of exciton condensation.\cite{safaei_quantum_2018} This theoretical metric has been used to explore the potential for exciton condensation in a variety of systems outside of the traditional regime of two-dimensional materials, including molecular layered systems\cite{safaei_quantum_2018,schouten_exciton_2021,sager_beginnings_2022} and amorphous materials.\cite{schouten_potential_2023}

This perspective explores recent progress towards understanding and realizing exciton condensation, with a particular focus on the development and application of a characteristic theoretical metric for exciton condensation, an eigenvalue greater than one in the particle-hole reduced density matrix. 
We begin by giving a brief overview of the theoretical foundations of exciton condensation, as well as experimental progress towards its realization. We then detail  insights gained through the exploration of ODLRO in excitonic systems to identify possible candidates for exciton condensation. These works have led to better understanding  of structure-property and geometric considerations for the rational design of exciton condensate materials. Further, applications of the eigenvalue metric have found that a ``critical seed" of exciton condensation manifests in molecular-scale fragments of exciton condensate systems, providing insight into how local short-range strongly correlated effects may give rise to macroscopic exciton condensation.\cite{sager_beginnings_2022,payne_torres_molecular_2024} The large eigenvalue metric for exciton condensation has also been found in molecular and amorphous systems, suggesting that the chemical space for exciton condensate materials may be much broader than previously thought.\cite{safaei_quantum_2018, schouten_exciton_2021,schouten_potential_2023} These studies have shaped a chemical perspective of exciton condensation, which has allowed researchers to establish a link between exciton condensation and the highly efficient energy transfer found in photosynthetic light-harvesting complexes, suggesting possible avenues for highly efficient energy transfer under reasonable conditions.\cite{schouten_exciton-condensate-like_2023} We conclude this perspective with a discussion of the fermion-exciton condensate, a novel condensate state containing an exciton condensate that has been prepared on a quantum device and characterized via off-diagonal long-range order.\cite{sager_potential_2020,sager_simultaneous_2022,sager_entangled_2022}

\section{Off-diagonal long-range order and superfluidity}
\label{ODLRO}
Central to the theoretical understanding of quantum condensation phenomena is the concept of off-diagonal long range order (ODLRO), which ties the emergence of condensation and superfluidity to long range correlation exhibited by the reduced density matrix (RDM).\cite{yang_concept_1962,coleman_structure_1963,sasaki_eigenvalues_1965,penrose_bose-einstein_1956} In the case of Bose-Einstein condensation, condensation of bosons results in the presence of off-diagonal elements in the one-boson RDM given by
\begin{equation}
^1D^{i}_{j} = \bra{\Psi}   \opdag{a}_i\op{a}_j\ket{\Psi}
\label{eq:1}
\end{equation}
where $\ket{\Psi}$ is the $N$-electron wavefunction and $\opdag{a}_i$ and $\op{a}_i$ are the fermionic creation and annihilation operators. \cite{penrose_bose-einstein_1956} The eigenvalues of the one-particle RDM correspond to the single-particle occupation numbers of the eigenstates. Eigenvalues greater than one correspond to natural orbitals occupied by more than one boson and hence, by definition, indicate the presence and extent of a superfluid Bose-Einstein condensate.\cite{penrose_bose-einstein_1956,pitaevskij_bose-einstein_2016}

The signature of ODLRO in quasibosonic paired fermions, and thus of fermion pair condensation, is an eigenvalue larger than one in the particle-particle reduced density matrix, whose elements are given in second quantization by
\begin{equation}
^2D^{i,j}_{k,l} = \bra{\Psi}   \opdag{a}_i\opdag{a}_j\op{a}_l\op{a}_k \ket{\Psi},
\label{eq:2}
\end{equation}
\noindent The eigenvalues of this matrix correspond to the occupation of fermion-fermion pair states such as electron-electron Cooper pairs.\cite{yang_concept_1962,sasaki_eigenvalues_1965} An eigenvalue greater than one in the particle-particle density matrix is thus indicative of more than one fermion-fermion pair occupying a single geminal (particle-particle quantum state) and hence, the condensation of the quasibosonic fermion pairs such as in BCS superconductivity.

Like the loosely-bound Cooper pairs of electrons, excitons are quasibosonic, making an analogous quasibosonic condensation of multiple particle-hole pairs into a single quantum state possible. Early work by Blatt, B\"oer, and Brandt,\cite{blatt_bose-einstein_1962} as well as Keldysh and Kopaev,\cite{sadovskii_possible_2023} proposed the possibility for exciton condensation. Kohn and Sherrington argued that condensates of excitons would be incapable of superfluidity because the condensate would be incapable of transferring charge or mass.\cite{kohn_two_1970} Kohn believed that the absence of ODLRO in the particle-particle RDM of an exciton system made superfluidity impossible for excitons, as opposed to the presence of ODLRO in the 2-RDM of superconducting Cooper-pair condensates.
However, contemporaries such as Luonid Keldysh, Eiichi Hanamura, and Hartmut Haug argued that an exciton condensate could be a superfluid because it was capable of the superfluid transfer of excitation energy.\cite{keldysh_coherent_2017,hanamura_will_1974} Despite Kohn's claims,
other early work showed that, ODLRO in an exciton system could exist in the \textit{particle-hole} RDM,\cite{garrod_particle-hole_1969,kohn_two_1970,haug_derivation_1975} However, the significance of ODLRO in the particle-hole RDM as the necessary condition for exciton condensation was not fully understood until the 2018 work of Safaei and Mazziotti.\cite{safaei_quantum_2018}

As detailed in the work of Safaei and Mazziotti, exciton condensation emerges from the entanglement of excitons, which results in off-diagonal long-range order in the particle-hole reduced density matrix, whose elements are given by
\begin{equation}
^2G^{i,j}_{k,l} = \bra{\Psi}   \opdag{a}_i\op{a}_j\opdag{a}_l\op{a}_k \ket{\Psi}.
\label{eq:3}
\end{equation}
There is an extraneous eigenvalue of the particle-hole RDM which corresponds to a ground-state-to-ground-state projection, rather than exciton condensation. This extraneous eigenvalue is removed by constructing a modified particle-hole matrix with the elements corresponding to the ground-state-to-ground-state projection removed. This modified particle-hole RDM is referred to as $^2\Tilde{G}^{i,j}_{k,l}$:
\begin{equation}
^2\Tilde{G}^{i,j}_{k,l}={}^2G^{i,j}_{k,l}-{}^{1}D^{i}_{j}{}^{1}D^{l}_{k},
\label{eq:4}
\end{equation}
where ${}^{1}D^{i}_{j}$ is the one-particle RDM. The eigenvalues and eigenvectors of this modified particle-hole RDM can then be calculated as
\begin{equation}
^2\Tilde{G}\nu_i=\lambda_i\nu_i.
\label{eq:5}
\end{equation}

For an uncorrelated system, the eigenvalues of the modified particle-hole RDM can be no greater than one. However, in a strongly-correlated system such as an exciton condensate, off-diagonal long-range order is present in the particle-hole RDM, resulting in an eigenvalue greater than one. The presence of an eigenvalue greater than one indicates multiple excitons populating a single particle-hole quantum state of the system and thus the emergence of exciton condensation. The magnitude of the largest eigenvalue gives the population of the lowest particle-hole quantum state and thus corresponds to the extent of exciton condensation.\cite{safaei_quantum_2018} This methodology allows for the investigation of molecular systems for $\lambda_G$ via the outputs from advanced electronic structure methods such as variational two-electron reduced density matrix theory (V2RDM)\cite{mazziotti_reduced-density-matrix_2007,mazziotti_contracted_1998,nakata_variational_2001,mazziotti_realization_2004,gidofalvi_active-space_2008,mazziotti_two-electron_2012,mazziotti_pure-_2016,mazziotti_large-scale_2011,cioslowski_many-electron_2000,zhao_reduced_2004,cances_electronic_2006,shenvi_active-space_2010,mazziotti_structure_2012,verstichel_variational_2012,piris_global_2017} or complete active space self-consistent-field theory (CASSCF),\cite{siegbahn_comparison_1980,roos_complete_1980,siegbahn_complete_1981} or other methods capable of capturing strong electron correlation. The large eigenvalue, which we will reference hereafter as $\lambda_G$, has been used as the metric for exciton condensation in a variety of works that will be discussed later in the Perspective.\cite{safaei_quantum_2018,sager_beginnings_2022,schouten_exciton_2021,schouten_potential_2023,schouten_exciton-condensate-like_2023,sager_preparation_2020,sager_potential_2020,sager_simultaneous_2022,sager_entangled_2022}

The eigenvalue metric for exciton condensation can also be obtained from only the cumulant part of the particle-hole RDM,\cite{schouten_large_2022} which is the ``connected'' part of the RDM that cannot be written as a product of lower order RDMs.\cite{mazziotti_contracted_1998,mazziotti_approximate_1998} Much of the correlation information contained in the RDM does not require the complete 2-RDM and can instead be obtained from the cumulant part alone. The signature of condensation from the cumulant also has the benefit of being size extensive, which could make it ideal for modeling scenarios in which the size of the condensate changes. Additionally, recent work has utilized the eigenvalue of the cumulant as an entanglement witness, with the elements of the cumulant obtained from solid-state spectroscopy, suggesting potential applications for the cumulant eigenvalue metric for exciton condensation in experimental contexts.\cite{liu_entanglement_2024}

Visualization of the theoretically calculated excitonic state can create a physical representation of the particle-hole pairing allowing for observation of excitonic characteristics such as delocalization and inter- or intralayer character. These characteristics are particularly relevant, as delocalized interlayer excitons should have greater exciton stability, which is desirable for exciton condensation. One common method for visualizing excitons is via the square modulus of the wavefunction obtained from solving the Bethe-Salpeter equation,\cite{salpeter_relativistic_1951} which is used to obtain the probability density of the excitonic electron corresponding to a fixed location of hole.\cite{sun_evidence_2022,ulman_organic-2d_2021,ataei_evidence_2021,varsano_monolayer_2020,jiang_spin-triplet_2020} Such visualization can obtained from a molecular structure via the outputs from advanced electronic structure methods such as V2RDM through the eigenvector or the modified particle-hole RDM corresponding to $\lambda_G$ and a matrix of the molecular orbitals in terms of atomic orbitals. For a more detailed description of this methodology, which is used in Refs. \citenum{safaei_quantum_2018, schouten_exciton_2021, sager_beginnings_2022, schouten_potential_2023, payne_torres_molecular_2024}, see the methods section of any of these works.


\section{Exciton condensate or excitonic insulator?}

\begin{figure}
\centering
\begin{subfigure}{0.49\linewidth}
\caption{}
\label{fig:phases_a}
    \includegraphics[width=\linewidth, height=0.25\textheight, keepaspectratio]{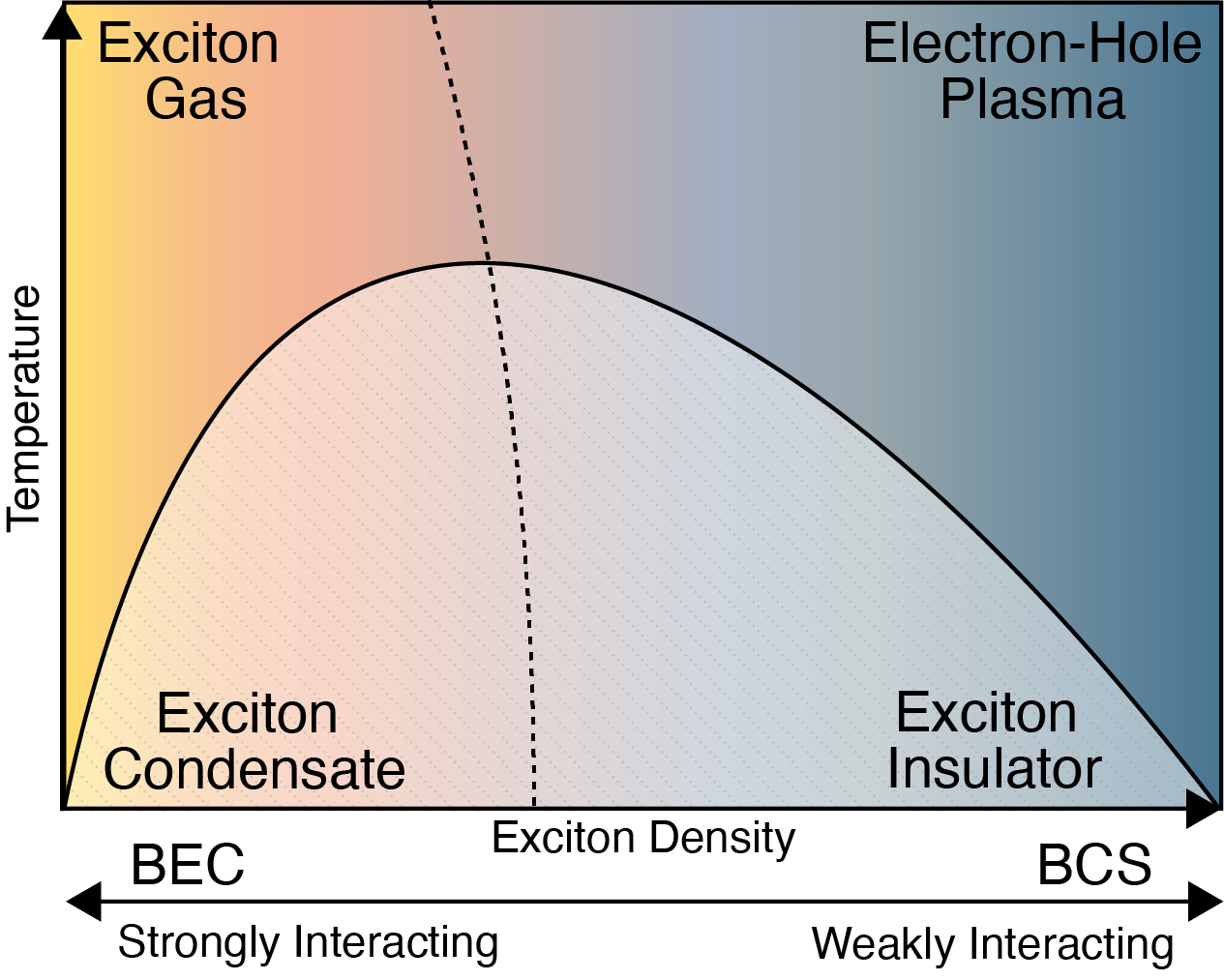}
\end{subfigure}
\hfill
\begin{subfigure}{0.49\linewidth}
\caption{}
\label{fig:phases_b}
    \includegraphics[width=\linewidth, height=0.25\textheight, keepaspectratio]{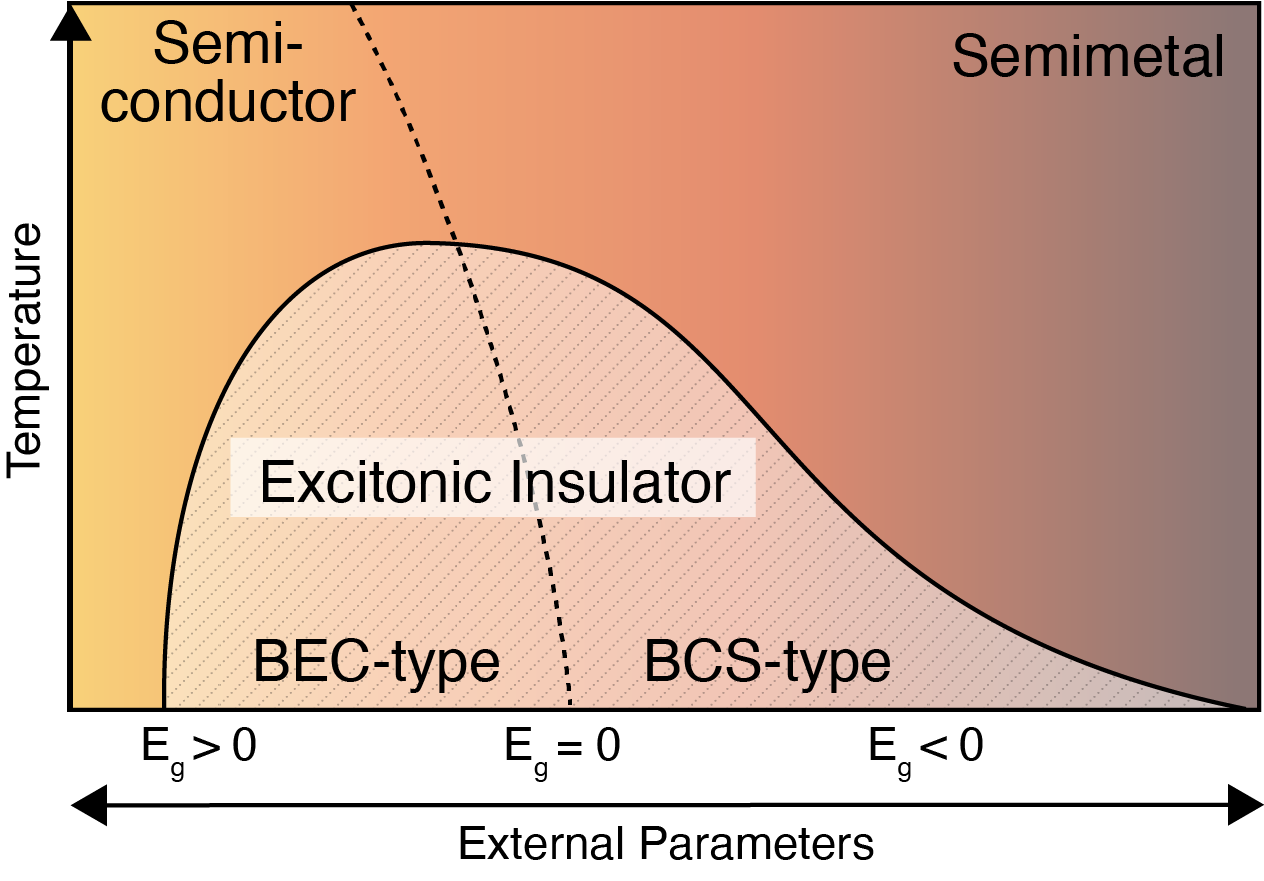}
\end{subfigure}
\caption{(a) Graphic representing the excitonic phases as defined by Refs. \cite{comte_exciton_1982} and \cite{nozieres_exciton_1982}. Adapted from Ref.\ \citenum{sager-smith_exploration_2023}. (b) Graphic representing the excitonic phases as defined by Ref.\ from Ref. \citenum{jerome_excitonic_1967}. Adapted from Ref.\ \citenum{phan_spectral_2010}. }
\label{fig:exciton_phases}
\end{figure}

Falling under the umbrella of exciton condensation are several strongly correlated excitonic phases, which are often discussed in different portions of the literature. Terminological differences also exist between experimentalists and theorists and between the different fields interested in exciton condensation. Definitions and terminology established by Comte and Noziéres \cite{comte_exciton_1982,nozieres_exciton_1982} classify the excitonic phases according to temperature and the density of the excitons. In the low-density limit, where the distance between electron-hole pairs is much larger than the radius of the electron-hole pair, the excitons are considered ``strongly interacting" and their behavior is well-described by Bose-Einstein statistics. Such condensates arise when an \textit{exciton gas}, a system of dilute excitons at high temperatures that behaves similarly to a Bose gas, is cooled until the excitons \textit{condense} into a single quantum state. This is likely to occur when $\sigma = m^*_e/m^*_h \approx 1$, where $m^*_e$ and $m^*_h$ are the electron and hole effective masses. When $\sigma \ll 1$, however, the exciton gas will tend to crystallize into a regular lattice of excitons. At high temperatures and densities, excitonic systems behave as an \textit{electron-hole plasma}, which when cooled forms either charge density waves or a condensate of diffuse, weakly interacting Cooper-pair-like excitons described by BCS theory. This high-density exciton condensate may also be termed an \textit{excitonic insulator}, so named because it is expected to be an insulating phase.\cite{kalt_excitonic_2024, hanamura_condensation_1977, sager-smith_exploration_2023, jerome_excitonic_1967, nozieres_exciton_1982, comte_exciton_1982} A crossover with a well-defined intermediate regime, rather than a phase transition, is expected between the BEC (exciton condensate) and BCS (exciton insulator) phases.\cite{comte_exciton_1982, liu_crossover_2022} A graphic representing these excitonic phases is provided in Fig.\ \ref{fig:phases_a}.

Other portions of the literature use the term excitonic insulator to describe a system in which the binding energy of excitons is greater than the band gap, thus making the system unstable to the spontaneous formation excitons that condense into the ground state.\cite{kalt_excitonic_2024, moskalenko_bose-einstein_2000} This state is referred to as an \textit{excitonic insulator} because it opens up the band gap of the system. In this context, exciton insulator may refer to either the exciton
condensate or exciton insulator or any intermediate phase described by Fig.\ \ref{fig:phases_a} as long as no excitation energy is supplied and the
excitons are indeed a ground state phenomenon. A graphic representing the excitonic phases as defined by the ground-state excitonic insulator terminology is provided in Fig.\ \ref{fig:phases_b}. Such an equilibrium exciton condensate state is thought to occur in narrow-gap semiconductors or semimetals. In this nomenclature, the BEC regime is defined as occurring in semiconductors (where the Coulomb interaction is essentially unscreened) and the BCS regime occurs in semimetals (where the Coulomb interactions are highly-screened).\cite{lu_zero-gap_2017} As in the case of the Comte-Noziéres framework described above, there is expected to be a smooth transition from the dilute BEC to dense BCS regimes.\cite{phan_spectral_2010}

In general, the various definitions of excitonic insulators and the BEC and BCS regimes may overlap, and terminological differences can lead to confusion between different areas of the literature. However, these various definitions can be united by the commonality of off-diagonal long-range order between them. Although structural or electronic differences may lead researchers to think of these various condensed excitonic phases separately, it is important to note that both exciton condensates and excitonic insulators (from either definition) involve the condensation of excitons into the ground state, and so will display off-diagonal long-range order in the modified particle-hole reduced density matrix. In the following work, exciton condensation will be used as a general term for any case in which off-diagonal long-range order is found in the modified particle-hole reduced density matrix, as this always describes the condensation of excitons.

\section{Counterflow Superconductivity}
\begin{figure}[ht]
    \centering
    \includegraphics[width=0.55\textwidth]{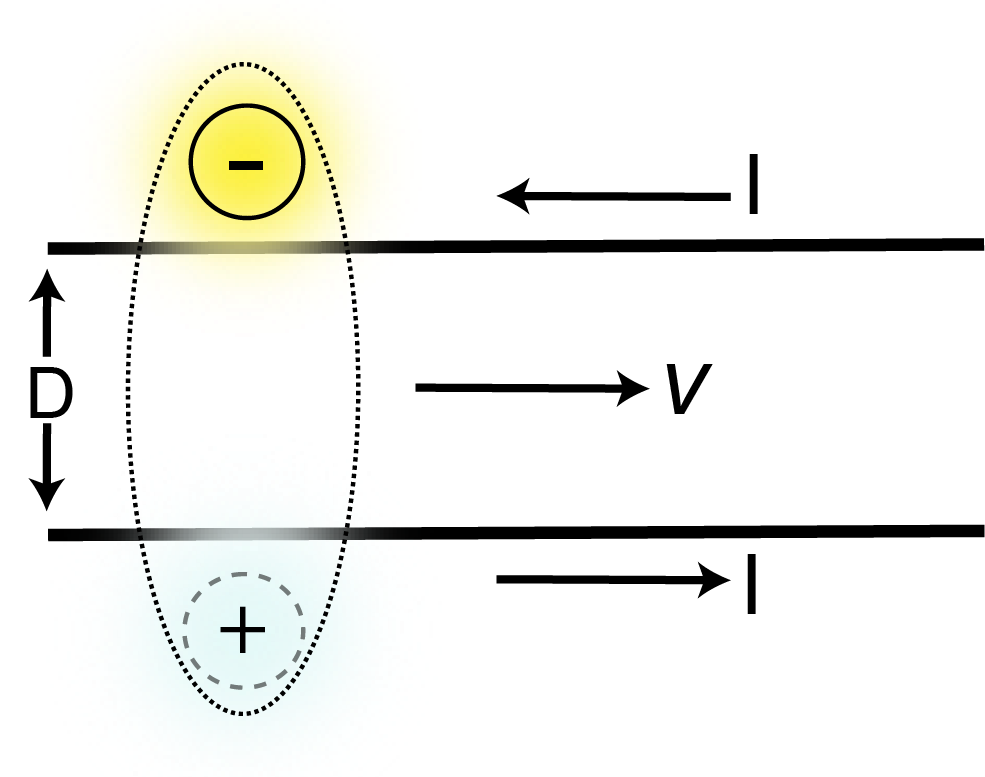}
    \caption{Proposed schematic diagram for achieving counterflow superconductivity, wherein a condensate of spatially indirect excitons in a bilayer system causes nonattenuating currents to flow in opposite directions in each layer. Adapted from Ref.\ \citenum{lozovik_new_1976} }
    \label{fig:counterflow_superconductivity}
\end{figure}

One of the most compelling aspects of exciton condensation is the idea that superfluid excitons in bilayers consisting of spatially separated electrons and holes could lead to a new kind of superconductivity: a \textit{counterflow} superconductivity where the superfluid flow of interlayer excitons results in nondissipative currents flowing in opposite directions in each layer.\cite{lozovik_new_1976} A schematic depicting counterflow superfluidity is shown in Fig.\ \ref{fig:counterflow_superconductivity}. The idea of counterflow superconductivity forms part of the inspiration for experimental efforts to realize exciton condensation in bilayer systems.

\section{Physical manifestations of exciton condensation}

Physical manifestation of exciton condensation could lead to many exciting technological advances, including ultra-efficient energy transport, novel electronic devices, and quantum computing. Much early work on the experimental realization of exciton condensation focused on bulk semiconductors such as Cu$_2$O and CuCl, which were believed to have sufficiently stable excitons at room temperature, as well as repulsive interactions between excitons that made competing excitonic phases such as biexcitons unlikely.\cite{snoke_boseeinstein_2014, ohara_strong_1999, snoke_quantum_1987,snoke_evidence_1990,lin_bose-einstein_1993,hulin_evidence_1980} However, experiments with CuCl were limited by short radiative lifetimes caused by coupling of photons to excitons, and experiments with Cu$_2$O were stymied by strong Auger recombination processes limiting the lifetimes of the excitons.\cite{snoke_boseeinstein_2014,wolfe_search_2014} Research then began to turn towards bilayer systems.\cite{sugakov_exciton_2006, eisenstein_boseeinstein_2004} Bilayer systems with spatially separated electrons and holes also have a number of properties that facilitate the formation of an exciton condensate: the spatial separation of electron and hole provides a barrier to recombination that extends exciton lifetimes, while the orientation of dipoles causes intralayer Coulombic repulsion which inhibits the formation of competing states such as biexcitons or electron-hole plasmas.\cite{fil_electron-hole_2018,eisenstein_boseeinstein_2004,eisenstein_exciton_2014} In particular, researchers found success in observing exciton condensation with indirect excitons in GaAs-based coupled quantum wells (CQWs),\cite{butov_condensation_1994,butov_exciton_2003,larionov_condensation_2001,krivolapchuk_specific_2001,butov_macroscopically_2002,high_spontaneous_2012} and in bilayer systems in the quantum hall regime.\cite{eisenstein_exciton_2014, kellogg_vanishing_2004, tutuc_counterflow_2004, spielman_resonantly_2000} Graphene bilayers have been of particular interest, as they afford the desired spatially indirect exciton, and the twist-angle dependence of their excitonic states gives a potential avenue for tuning the system towards exciton condensation.\cite{hu_quantum-metric-enabled_2022, rickhaus_correlated_2021, sager_beginnings_2022} Several experiments have demonstrated strong evidence for exciton condensation in graphene bilayer systems.\cite{rickhaus_correlated_2021, liu_quantum_2017, li_excitonic_2017, li_pairing_2019, burg_strongly_2018, li_negative_2016} However, such experiments often require very high magnetic fields, limiting the practical applicability of these systems. Another promising platform for the realization of exciton condensation are van der Waals heterostructures composed of two-dimensional transition metal dichalcogenide (MX$_2$: M = Mo, W; X = S, Se, Te) bilayers. Such excitons have a large binding energies around 0.5 eV\cite{mak_photonics_2016} and are predicted to have maximum condensation temperatures in the realm of room temperature.\cite{fogler_high-temperature_2014, wu_theory_2015, berman_high-temperature_2016, debnath_exciton_2017} Interlayer excitons are favored due to indirect band gaps, facilitating the long exciton lifetimes required for exciton condensation.\cite{rivera_interlayer_2018} Recent experimental work has shown evidence for exciton condensation in MoSe$_2$-WSe$_2$ heterostructures.\cite{wang_evidence_2019, sigl_signatures_2020, ma_strongly_2021} Other systems of interest for exciton condensation include a wide variety of low-dimensional materials such as 2D 1T-TiSe$_2$,\cite{monney_probing_2010, cercellier_evidence_2007, kogar_signatures_2017} monolayer WTe$_2$,\cite{song_signatures_2023, sun_evidence_2022}, and Ta$_2$NiSe$_5$,\cite{lu_zero-gap_2017,werdehausen_coherent_2018,kim_direct_2021} which has a quasi-1D structure.

Various experimental signatures have been used to provide evidence for the creation of exciton condensates. In early experiments seeking to observe exciton condensation in bulk semiconductors such as Cu$_2$O, researchers looked for signs of exciton condensation in the phonon-assisted luminescence spectra, which gives the kinetic energy distribution of the excitons. These spectra were compared to fitted Bose-Einstein and Maxwell-Boltzmann distributions, and relative density data (from luminescence intensity and exciton cloud volume) were compared to densities calculated from fits to the Bose-Einstein distribution and the phase boundary for BEC.\cite{snoke_boseeinstein_2014, snoke_quantum_1987, snoke_evidence_1990, lin_bose-einstein_1993} However, these experiments had ambiguities resulting from the fact that the three-dimensional data must be integrated over at least one dimension.\cite{snoke_boseeinstein_2014} Other experiments focused on exciton transport, as excitons ceasing to scatter with acoustic phonons would indicate the onset of superfluidity for free excitons in a crystal. However, as of yet no experiment has conclusively shown exciton condensation in Cu$_2$O, as discussed in the previous paragraph.

As researchers turned to bilayers systems, new signatures of exciton condensation were employed. One such signature is the presence of so-called perfect Coulomb drag, wherein the strong interlayer Coulomb attraction between the electrons and holes in the bilayer means that a transport current driven through one layer of the bilayer system should induce an equal current of holes in the other layer.\cite{nandi_exciton_2012, su_how_2008} The quantum Hall effect affords another experimental signature of exciton condensation: vanishing Hall resistance for a counterflow experiment. In the quantum Hall effect, a current $I_x$ flowing in the $x$-direction (with drag $R_{xx}$) in a bilayer system subject to a magnetic field in the $z$-direction ($B$) will cause a voltage $V_H$ in the $y$ direction with a Hall resistance given by $R_{xy}=\frac{2\pi\hbar}{e^2\nu}$. The formation of an exciton condensate is evidenced by the vanishing of the longitudinal components of the drag (and the presence of a large $V_H$) when equal currents are made to flow in parallel in each layer. This vanishing hall resistance has been used as evidence for the formation of an exciton condensate in several bilayer systems.\cite{liu_quantum_2017, li_excitonic_2017, kellogg_observation_2002, kellogg_vanishing_2004} Other experiments in bilayer systems observe the interlayer tunneling of particles in the bilayer. Interlayer tunneling of electrons (or holes) in a bilayer system is maximized upon the formation of an exciton condensate due to the presence of correlated electron–hole pair tunneling---the recombination of bound electron–hole pairs.\cite{eisenstein_boseeinstein_2004, xie_electrical_2018} The maximized tunneling is characterized by the presence of dramatic peak in the plot of interlayer current ($dI/dV$) versus applied interlayer voltage ($V$) at zero voltage.\cite{eisenstein_boseeinstein_2004, spielman_resonantly_2000, tiemann_critical_2008, tiemann_dominant_2009} This characteristic has been used as evidence for exciton condensation in a variety of coupled quantum wells and quantum Hall bilayers.\cite{eisenstein_precursors_2019,jang_strong_2021,zhang_quasiparticle_2020,spielman_onset_2004} The maximum interlayer tunneling also corresponds to strong electroluminescence from the recombination of the interlayer excitons, which should be strongly dependent on exciton density. Such electroluminescence findings were recently used as evidence for the existence of  exciton condensation in MoSe$_2$-WSe$_2$ bilayers.\cite{wang_evidence_2019}

Aside from these approaches, other techniques employed recently to provide evidence for exciton condensation include those which probe the electronic structure of materials, in particular the band structure. Angle-resolved photoemission spectroscopy, which allows for the study of the electronic structure by probing the energies and momenta of the electrons in a material, has been used to observe the band mixing, band gap opening, and band flattening characteristic of exciton condensation.\cite{monney_probing_2010, gao_evidence_2023, song_signatures_2023, huang_evidence_2024} Scanning tunneling spectroscopy, which investigates the density of electrons in a material as a function of their energy, has been used to identify the characteristic opening of the band gap to identify a potential excitonic insulator in quantum-confined Sb nanoflakes,\cite{li_possible_2019} and to observe electronic features attributed to the formation of an excitonic insulator in Ta$_2$NiSe$_5$.\cite{lee_strong_2019} Raman spectroscopy has also been used to observe features of the excitonic phase transition.\cite{kim_direct_2021,ataei_evidence_2021} Momentum-resolved electron energy-loss spectroscopy (M-EELS) can be used to measure the collective excitations of a material, and has been used to demonstrate that the energy needed to excite an electronic mode in the three-dimensional semimetal 1\textit{T}-TiSe$_2$ becomes insignificant at a finite momentum, which researchers believe signifies the emergence of an exciton condensate.\cite{kogar_signatures_2017}

Experimental realization of exciton condensation is still a very active area of research, and new experimental methods and condensate candidates are continually being explored. For a more detailed discussion of the experimental techniques and signatures that have been used in the study of exciton condensation, please see Refs.\ \citenum{kalt_excitonic_2024} and \citenum{eisenstein_exciton_2014}.

\section{A ``critical seed" of exciton condensation}

Having given a brief overview of the broad field of exciton condensation, including theoretical foundation and experimental progress, we will now detail recent insights gained through using the eigenvalue metric for exciton condensation to study emergent exciton condensation and exciton-condensate-like phenomena.

\begin{figure}
\centering
\null\hfill
\begin{subfigure}[c]{0.39\textwidth}
\caption{}
\label{fig:criticalseed_a}
    \centering\includegraphics[height=0.35\textheight]{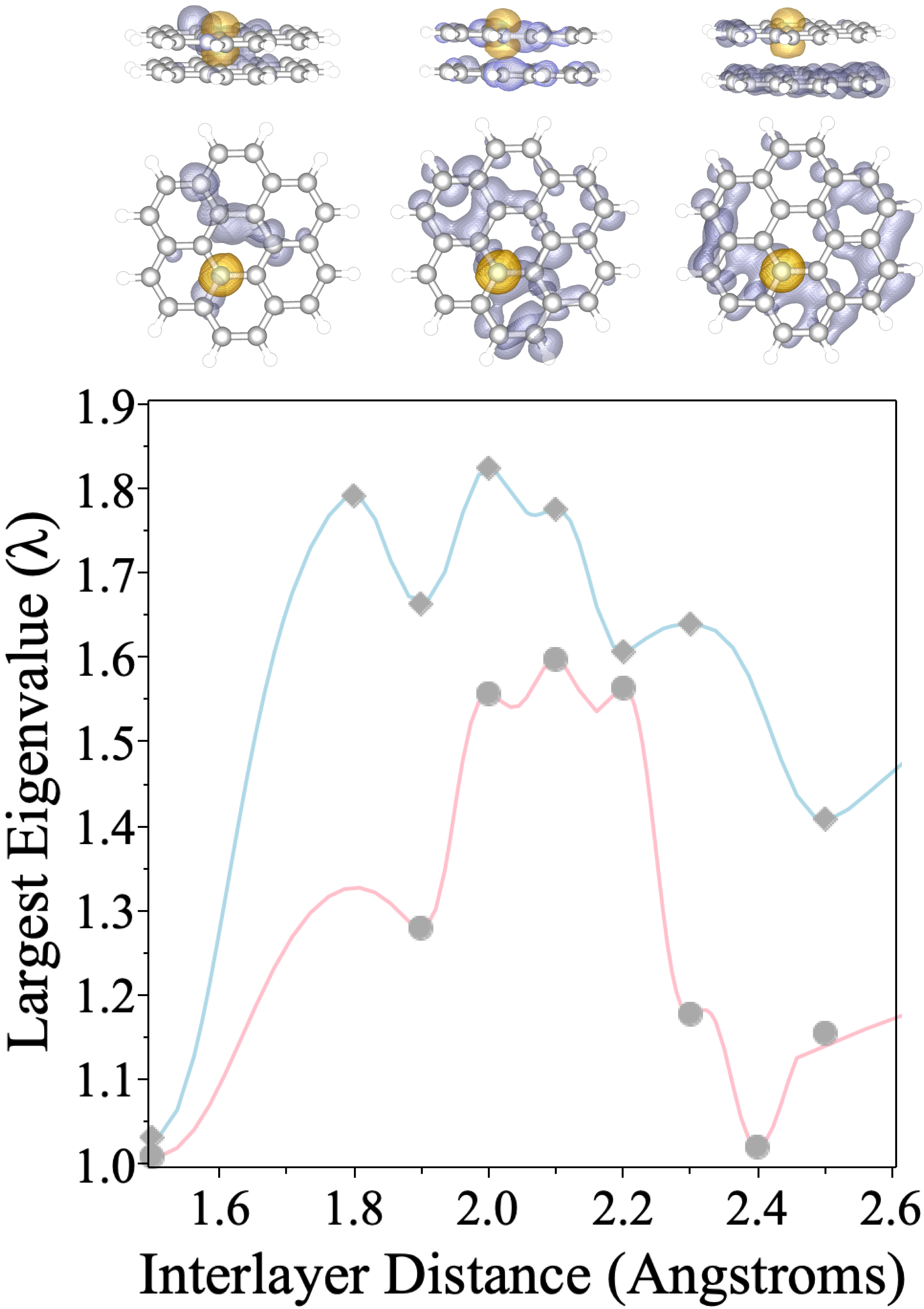}
\end{subfigure}
\null\hfill
\begin{subfigure}[c]{0.59\textwidth}
\caption{}
\label{fig:criticalseed_b}
    \centering\includegraphics[height=0.35\textheight]{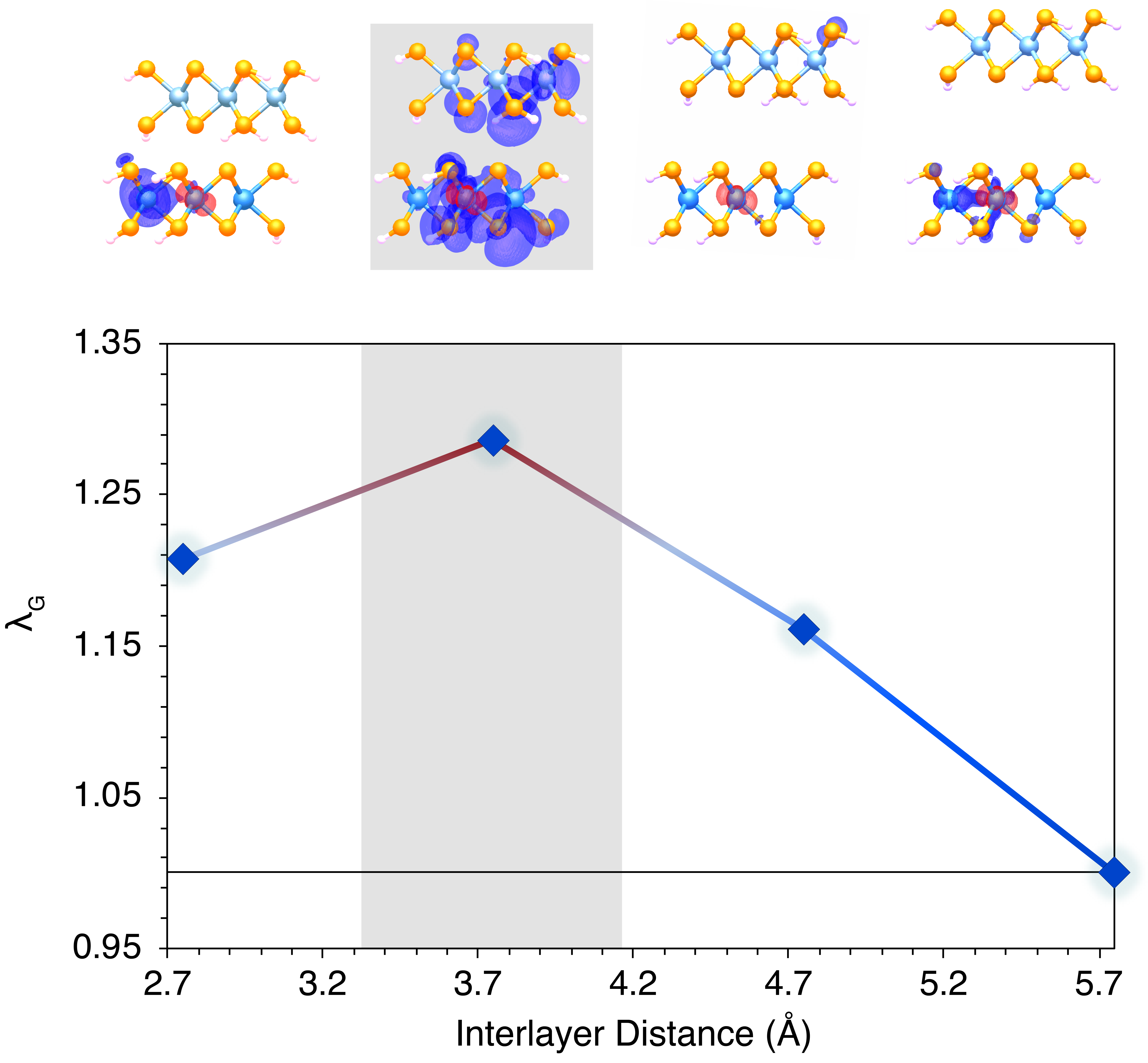}
\end{subfigure}
\null\hfill
\caption{(a) Top: Visualization of the exciton density for coronene bilayer systems with an inter-layer offset angle of $0\degree$, with inter-layer distances (from left to right) of 1.5 {\AA}, 2.0 {\AA}, and 2.5 {\AA}. The probablistic location of the hole for the exciton associated with the large eigenvalue is plotted in gray-violet, with respect to the fixed atomic orbital plotted in gold. Bottom: The exciton population in a single quantum state ($\lambda_G$) over various distances between coronene layers, from V2RDM-CASSCF calculations using a [24,24] active space (blue) and CI-CASSCF calculations using a [10,10] active space (pink). (c) Top: Visualization of the exciton density of the exciton mode corresponding to the largest eigenvalue of the particle-hole density matrix ($\lambda_G$) for interlayer spacings of MoSe$_2$-WSe$_2$ ranging from 3.745 to 5.745 {\AA}. The the probabilistic density of the electron for the exciton associated with the large eigenvalue is plotted in purple, while the atomic orbital corresponding to the constrained location of the hole is plotted in red. Bottom: The large eigenvalue $\lambda_G$ plotted against the distance between MoSe$_2$-WSe$_2$ layers, from V2RDM-CASSCF calculations using a [8,8] active space. Panel (a) top and bottom reprinted from Ref.\ \citenum{sager_beginnings_2022}, with the permission of AIP Publishing. Copyright 2022 AIP Publishing LLC. Panel (b) top and bottom reprinted with permission from Ref. \citenum{payne_torres_molecular_2024}. Copyright 2024 The Authors. }
\label{fig:critical_seed}
\end{figure}

The large eigenvalue of the modified particle-hole RDM has recently been used to show the \textit{beginnings} of exciton condensation in molecular scale fragments of materials that have shown experimental evidence for exciton condensation,\cite{safaei_quantum_2018, schouten_exciton_2021, sager_beginnings_2022, schouten_potential_2023,payne_torres_molecular_2024} including a coronene analog of a graphene double layer\cite{sager_beginnings_2022} and molecular fragments of several van der Waals heterostructure bilayers.\cite{payne_torres_molecular_2024} While the $\lambda_G$ values obtained in these studies are not truly macroscopic, all are greater than one, indicating more than one exciton populating the ground state. We believe that such eigenvalues instead demonstrate a ``critical seed" of condensation: the molecular origins of the condensation that would be observed (with a macroscopic eigenvalue) in material-scale samples of these systems.

Graphene double layers have shown experimental evidence for exciton condensation, resulting in near-dissipationless exciton transport.\cite{li_excitonic_2017,liu_quantum_2017} In these experiments the exciton condensate phase is formed in the quantum Hall regime, wherein tuning each layer such that the lowest Landau level is half-full corresponds to populating the lowest band in each layer with electrons and holes in equal number, which then bind to form a system of indirect excitons. Exciton condensate behavior was evinced by the quantization of the Hall resistance $R_{xy}$ (the quantum Hall effect) and a local zero value of $R_{xx}^{drag}$. Inspired by this experimental evidence of exciton condensation in graphene bilayers, Sager et\ al.\ sought to determine if they could observe the \textit{theoretical} metric for exciton condensation in a similar, but molecular-scaled, material.\cite{sager_beginnings_2022} The results showed that a coronene bilayer, wherein each coronene layer is a seven-benzene-ring patch of graphene, could display an eigenvalue greater than one. This large eigenvalue is tuneable with interlayer distance (as seen in Fig, \ref{fig:critical_seed}b), as well as interlayer twist-angle. Distances around 2.0 {\AA} are found to maximize condensation (with $\lambda_G = 1.824$), while the eigenvalue population drops off at both larger and smaller distances. Additionally, interlayer delocalization of the hole increases as the exciton condensate character increases. Maximal exciton condensate character is observed for an inter-layer offset angle of $0\degree$, but a large degree of condensation ($\lambda_G > 1.45$) is present for angles between $0\degree$ and $2\degree$, which are reminiscent of the angles used in magic-angle graphene studies.\cite{cao_unconventional_2018}

The beginnings of exciton condensation were also seen in a molecular scale fragment of MoSe$_2$-WSe$_2$, a van der Waals heterostructure that has recently shown experimental evidence for exciton condensation.\cite{wang_evidence_2019, sigl_signatures_2020, ma_strongly_2021} The potential exciton condensate phase in the experimental studies was evidenced by strong electroluminescence from the recombination of the interlayer excitons, corresponding to maximum interlayer tunneling (as discussed above). The theoretical work by Payne Torres et\ al.\ \citenum{payne_torres_molecular_2024} demonstrated the presence of an eigenvalue greater than one in a molecular scale fragment of MoSe$_2$-WSe$_2$. The degree of condensation increases with the size of the molecular fragment, indicating the possibility that the molecular fragment contains a ``critical seed" of the condensation that would be present in a bulk system. Further, the presence of a ``critical seed" suggests that local, short-range strongly correlated effects are significant for the formation of a macroscopic condensate. As in the case of the coronene bilayer, the extent of exciton condensation is also tuneable via interlayer distance and twist angle. Maximal exciton condensation occurs ($\lambda_G$ = 1.29) at an intermediate distance of 3.745 {\AA}, while increasing or decreasing the interlayer distance from that value diminishes the extend of condensation. Similar to the case of the coronene double layers, the interlayer delocalization of the exciton (represented in this case by the delocalization of the excitonic electron) is maximized for the calculation wherein $\lambda_G$ is also maximized. Probing the effect of inter-layer alignment angle via the six high symmetry stacking order of MoSe$_2$-WSe$_2$ (as defined by Ref.\ \citenum{zhao_excitons_2023}) reveals that the $R^h_h$ alignment (wherein each metal atom is aligned directly over a metal atom on the opposite layer and each chalcogen atom is aligned directly over a chalcogen atom on the opposite layer) results in maximized exciton populations around 1.5. This suggests that maximizing the alignment between sites on opposing layers of TMD bilayers may be significant for the emergence of large $\lambda_G$.

While it is known that ODLRO is the necessary condition for superfluidity, now the observation of the beginnings of exciton condensation in molecular-scale fragments of bilayer graphene and MoSe$_2$-WSe$_2$ connects the presence of ODLRO in the paricle-hole RDM to experimental results for exciton condensation. These results also give unique insight into the molecular origins of the phenomena that is observed in material-scale samples. While exciton condensation is generally understood as a macroscopic material-scale phenomenon, the presence of a ``critical seed" of condensation in a very small fragment of the material suggests that short-range strongly correlated effects on the molecular scale may be significant for the formation of a macroscopic condensate. Moreover, both the coronene double layer and MoSe$_2$-WSe$_2$ results highlight the importance of geometric considerations such as interlayer distance and twist angle in the design of exciton condensate materials. These findings are reminiscent of the magic-angle graphene studies, wherein interlayer twist-angle modulates the electronic structure of the material and impacts the appearance of unconventional superconductivity.\cite{cao_unconventional_2018} Additionally, the fact that interlayer delocalization of the exciton was maximized as the exciton condensate character was maximized both for coronene double layers and MoSe$_2$-WSe$_2$ suggests that delocalization of the exciton between layers is a strong indicator of high degrees of exciton condensation, and may in fact be necessary for a condensate to form. These results also suggest that exciton condensation may also be possible in molecular scale systems related to the extended systems that are of interest for exciton condensation, providing an exciting avenue for future study. Due to the presence of the eigenvalue metric for exciton condensation, it is possible that such systems may demonstrate novel counterflow or electroluminescence properties similar to those found in their extended counterparts.

\subsection{Molecular and amorphous exciton condensation}

\begin{figure}
\centering
\hfill
\begin{subfigure}{0.32\linewidth}
\caption{}
\label{fig:molecular_a}
    \includegraphics[height=0.18\textheight, keepaspectratio]{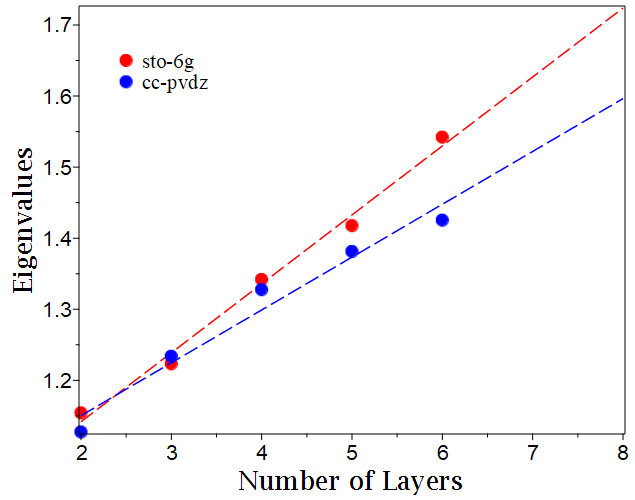}
\end{subfigure}
\hfill
\begin{subfigure}{0.18\linewidth}
\caption{}
\label{fig:molecular_b}
    \centering\includegraphics[height=0.18\textheight, keepaspectratio]{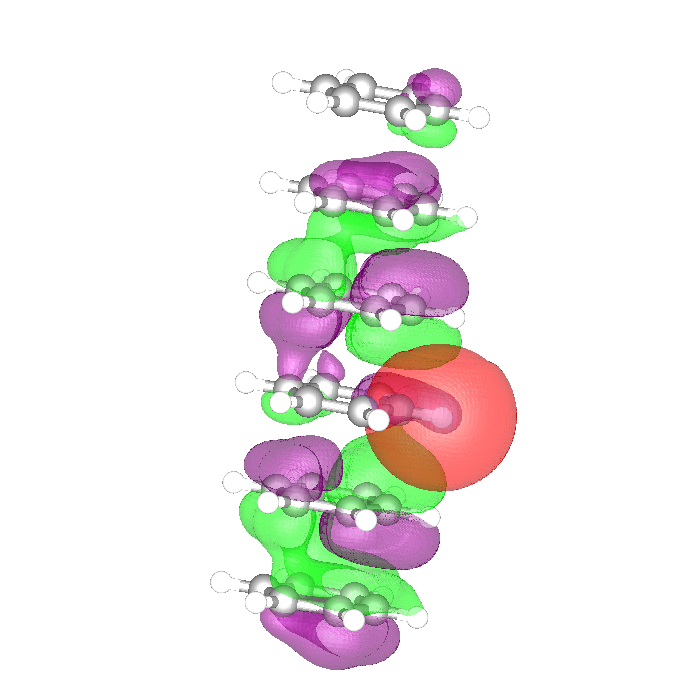}
\end{subfigure}
\hfill
\begin{subfigure}{0.29\linewidth}
\caption{}
\label{fig:molecular_c}
    \centering\includegraphics[height=0.18\textheight, keepaspectratio]{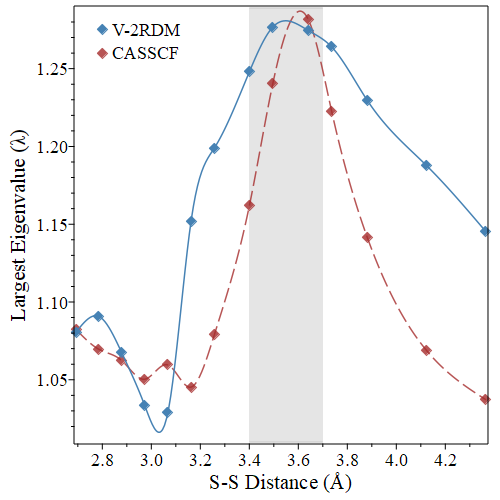}
\end{subfigure}
\hfill
\begin{subfigure}{0.18\linewidth}
\caption{}
\label{fig:molecular_d}
    \centering\includegraphics[height=0.18\textheight, keepaspectratio]{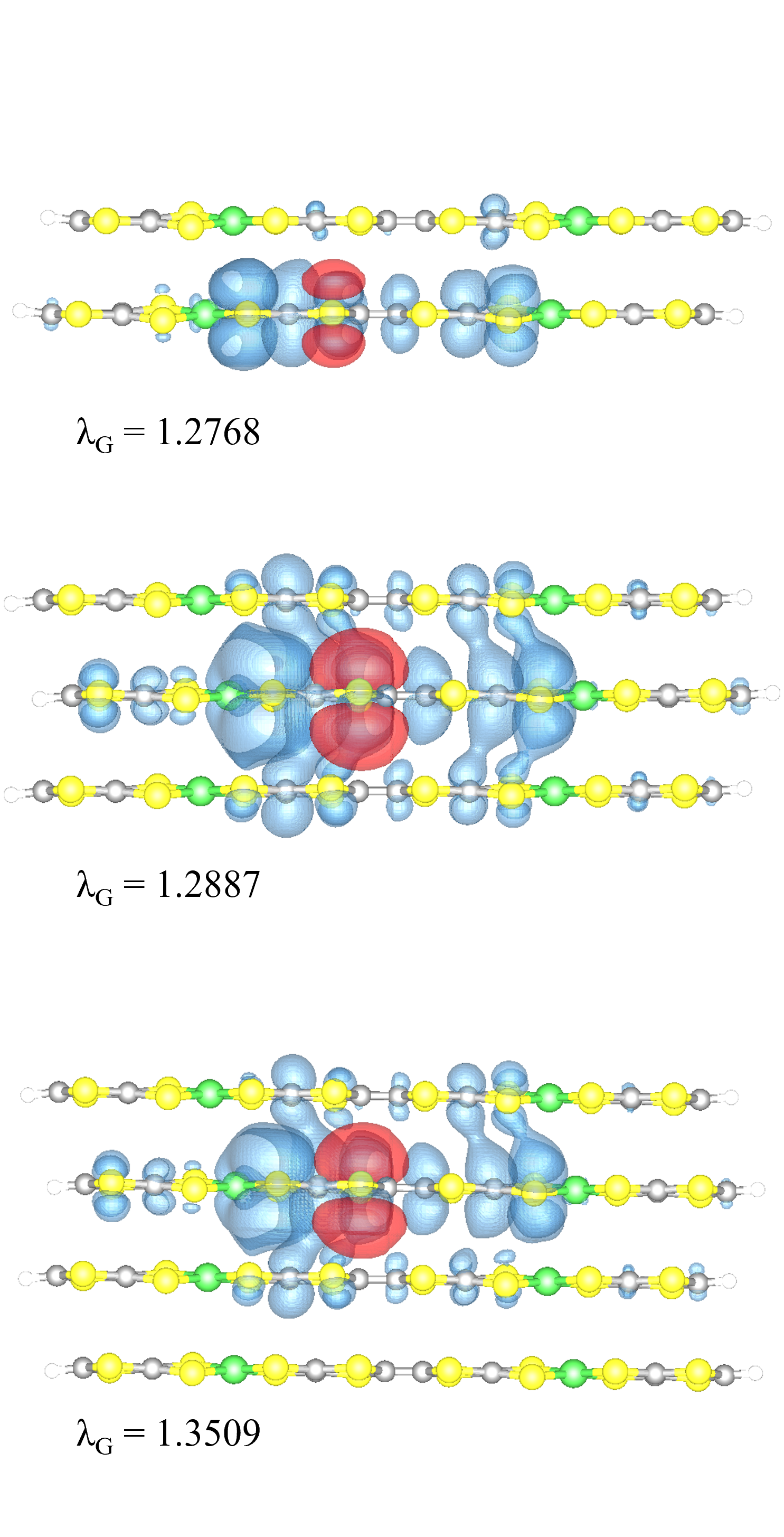}
\end{subfigure}
\hfill
\caption{(a) Plot of the largest eigenvalue of the particle-hole RDM versus the number of layers of benzene molecules from V2RDM-CASSCF calculations with the sto-6g (red) and cc-pvdz (blue) basis sets (all interlayer distances set to 2.5 {\AA}). (b) Visualization of the exciton density of the exciton mode corresponding to the largest eigenvalue of the particle-hole density matrix ($\lambda_G$) for a system of six layered benzene molecules, with the atomic orbital representing the constrained location of the electron plotted in red and the probabilistic density of corresponding hole plotted in purple and green. (c) Plot of $\lambda_G$ (blue) and the largest eigenvalues of the particle-particle density matrix (red) versus the interlayer distance for a bilayer of NiTTFtt molecules with a parallel offset of 1 {\AA}. (d) Visualization of the exciton density of the exciton mode corresponding to the largest eigenvalue of the particle-hole density matrix ($\lambda_G$), with the atomic orbital representing the constrained location of the electron plotted in red and the probabilistic density of corresponding hole plotted in blue for NiTTF van der Waals systems composed of two, three, and four layers. Panels a and b reprinted with permission from Ref.\ \citenum{schouten_exciton_2021}. Copyright 2021 The Authors, published by American Chemical Society. Panels  c and d reprinted with permission from Ref.\ \citenum{schouten_potential_2023}. Copyright 2023 American Physical Society.}
\label{fig:molecular_condensate}
\end{figure}

In addition to revealing the beginnings of exciton condensation in molecular-scale analogues of materials that have shown experimental evidence of exciton condensation, the large eigenvalue has also been used to show the potential for exciton condensation in several novel molecular and amorphous material systems that had not previously been considered for exciton condensation. Inspired by recent revelation that exciton condensation may be possible in systems other than extended materials,\cite{safaei_quantum_2018} Schouten et\ al.\ probed vertical stacks of benzene molecules for the presence of $\lambda_G$ greater than one.\cite{schouten_exciton_2021} Stacked benzene molecules have strong interlayer van der Waals interactions similar to those found in exciton condensate materials like graphene and van der Waals heterostructure bilayers, as well as $\pi-\pi$ interactions which facilitate electron transport between layers.\cite{solomon_chameleonic_2010, schneebeli_single-molecule_2011, wu_pi-conjugated_2010} The study found that exciton condensate character emerges for stacks of two and greater benzene molecules with spacings between 1.75 and 3 {\AA}. Exciton condensate character increases as the number of benzene molecules increased, as shown in Fig.\ \ref{fig:molecular_a}. Exciton condensate character is maximized when the benzene molecules have an interlayer alignment angle of $0\degree$, similar to the result seen in the study of coronene double layers discussed above. Comparing the benzene stacks to stacks of substituted benzene molecules aniline and fluorobenzene demonstrates that the addition of electron withdrawing or donating groups can reduce the extent of exciton condensation.

The potential for exciton condensation has also been found in an amorphous nickel tetrathiafulvalene-tetrathiolate (NiTTFtt) polymer.\cite{schouten_potential_2023}  NiTTFtt was thought to be a promising candidate for a potential exciton condensate because it exhibits $\pi-\pi$ overlap, intermolecular S-S interactions, and delocalization along the molecular chain.\cite{kobayashi_single-component_2004, xie_intrinsic_2022} Such properties have the potential to promote the stable, delocalized interlayer excitons desirable for exciton condensation. Probing NiTTFtt for $\lambda_G$ revealed an eigenvalue greater than one, with the amount of exciton condensate character dependent on the interlayer spacing, and increasing with the number of layers used in the truncated model system (\ref{fig:counterflow_superconductivity}c-d). The degree of exciton condensation is also increased in NiTTFtt dimers of longer chain length and in a model crystal structure fragment that introduces horizontal stacking. These results together suggest that the mechanism of the exciton condensate behavior depends on $\pi-\pi$ stacking, chain length of NiTTFtt dimers, and side-to-side S-S interactions. Shifting dimers of NiTTFtt parallel of perpendicularly to mimic the amorphous structure increased $\lambda_G$, suggesting that the amorphous nature of NiTTFtt may lead to enhanced exciton condensate character over an ordered structure.

The observation of the beginnings of exciton condensation both in molecular stacks of benzene molecules and in a molecular-scale model of an amorphous polymer suggests that exciton condensation may be possible in unique systems very different from the ordered 2D materials commonly researched. Furthermore, we have gained insight into the structure-property relationships that may impact the mechanism of exciton condensate formation in the molecular and amorphous systems.  The mechanism of the exciton condensate behavior in both cases appears to depend on $\pi-\pi$ stacking, evidenced by the impact of layer spacing and the number of layers on the degree of exciton condensation. As such, similar stacked materials with strong interlayer $\pi-\pi$ interactions are of interest for investigation into the potential for exciton condensation. The fact that chemical functionalization impacted the degree of exciton condensation in the benzene stacks suggests that chemical functionalization may be a route for maximizing the exciton condensation in molecular systems. As electron donating or withdrawing subunits create an imbalance of electrons and holes (reducing the number of electron-hole pairs available for condensation), this idea is similar to experimental literature which suggests that maintaining a balance of electrons and holes is an important factor for condensation.\cite{liu_quantum_2017} Moreover, we believe that the results discussed here motivate experimental investigations into exciton condensation in bulk samples of benzene stacks and NiTTFtt materials. Because the presence of ODLRO is indicative of superfluidity, experimental studies into these molecular exciton condensates may reveal similar novel conductivity behavior (such as ideal Coulomb drag or vanishing Hall resistance) similar to the behavior seen in bilayer exciton condensates. While it is still unknown if the coherent state of excitons that could be prepared in such molecular systems would share \textit{all} of the characteristics expected in the case of a traditional material-scale exciton condensate, the prospect of molecular long-range coherence is still exciting, and has the potential to lead to innovations in materials for highly efficient energy transport.

\section{Exciton-condensate-like features in photosynthetic light harvesting}
Observing the beginnings of exciton condensation in small molecular-scale systems inspired Schouten et\ al.\ to look for the eigenvalue metric for exciton condensation in small-scale molecular systems involved in excitonic processes.\cite{schouten_exciton-condensate-like_2023} Besides exciton condensate materials, another example of systems that exhibit excitonic coherence are photosynthetic light harvesting complexes that transfer energy via photogenerated excitons,\cite{engel_evidence_2007} shuttling excitation energy to a photosynthetic reaction center through a series of chromophores.\cite{adolphs_how_2006} While the study of exciton condensation has generally been completely unconnected from the study of energy transfer in light harvesting complexes, both phenomena involve the highly efficient transfer of energy via excitons. Additionally, a large body of work has been dedicated to exploring the impact of quantum effects such as strong correlation,\cite{mazziotti_effect_2012, davidson_eliminating_2022, mattioni_design_2021} dephasing,\cite{plenio_dephasing-assisted_2008, caruso_highly_2009, forgy_relations_2014} quantum coherence,\cite{engel_evidence_2007,hu_dark_2018,zerah_harush_photosynthetic_2021,tomasi_coherent_2019,collini_coherently_2010} and entanglement \cite{saberi_energy_2016,sarovar_quantum_2010,skolnik_cumulant_2013,fassioli_distribution_2010,dutta_delocalization_2019} in photosynthetic light-harvesting complexes. 
\begin{figure}
\centering
\hfill
\begin{subfigure}{0.32\linewidth}
\caption{}
\label{fig:FMO_a}
    \centering\includegraphics[height=0.2\textheight, keepaspectratio]{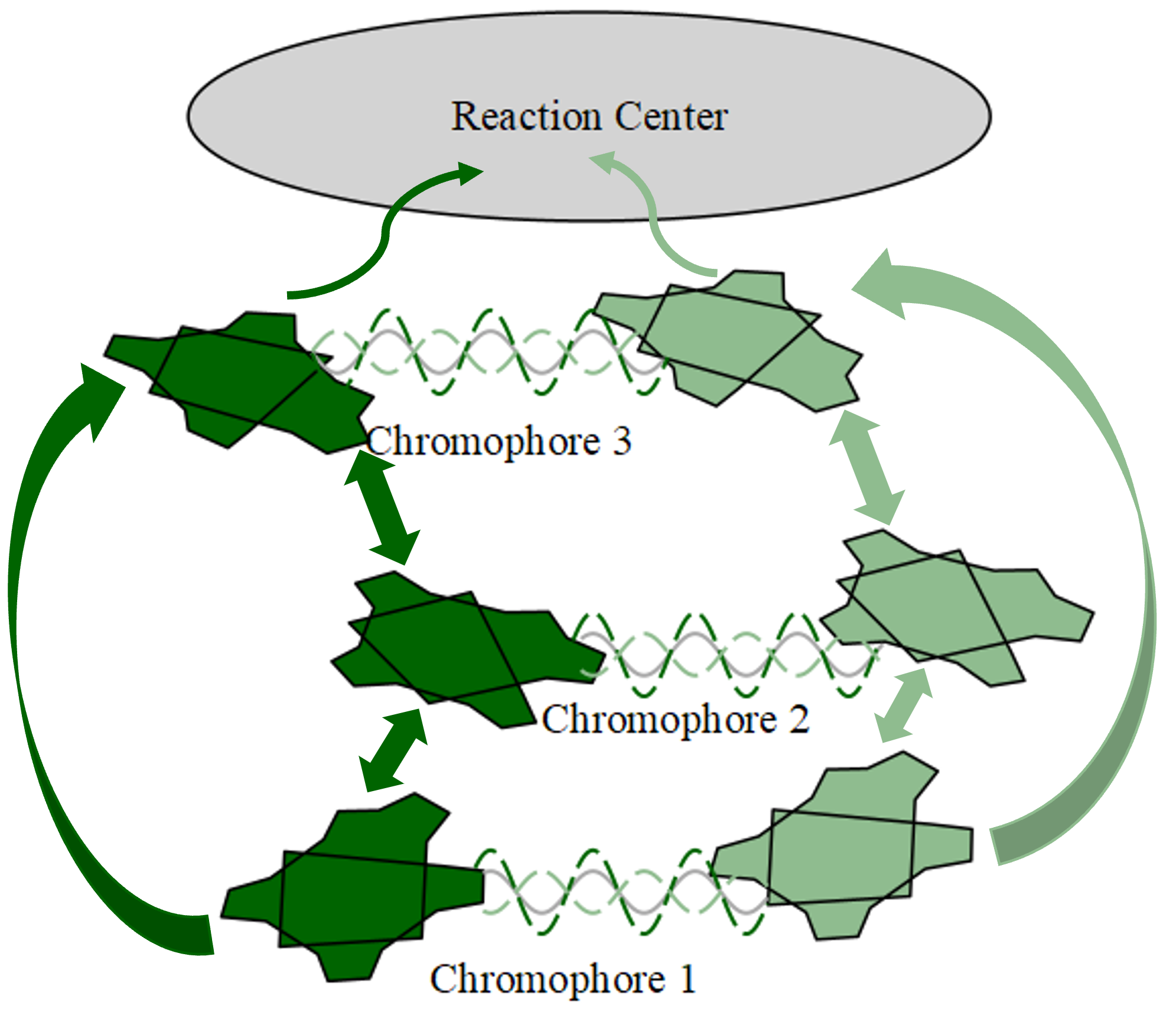}
\end{subfigure}
\hfill
\begin{subfigure}{0.29\linewidth}
\caption{}
\label{fig:FMO_b}
    \centering\includegraphics[height=0.2\textheight, keepaspectratio]{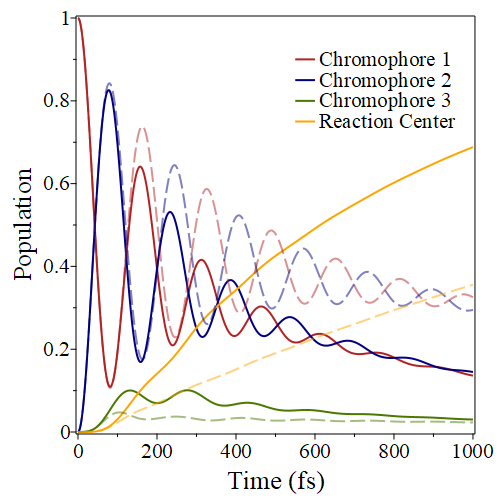}
\end{subfigure}
\hfill
\begin{subfigure}{0.29\linewidth}
\caption{}
\label{fig:FMO_c}
    \centering\includegraphics[height=0.2\textheight, keepaspectratio]{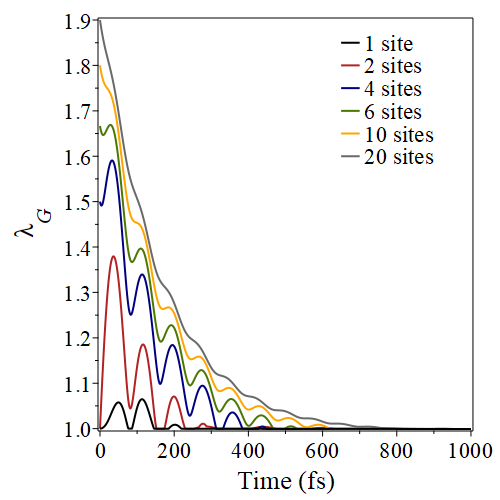}
\end{subfigure}
\hfill
\caption{(a) Two pathways for exciton transfer represented for two sites per chromophore, with no \textit{inter}chromophore cross-site coupling. When $V>0$, there is coupling between sites on the same chromophore, resulting in quantum interference between the sites as depicted. The quantum interference can be either constructive or destructive, with constructive interference enhancing the energy-transfer efficiency. (b) Population dynamics of the one-body model of the FMO complex (pale dashed lines) and the coupled model of the FMO complex with V = 0.6 and M = 20 (dark solid lines). (c) Dynamics of $\lambda_G$ for the FMO complex with M = 1, 2, 4, 6, 10, and 12 with an entangled initial excitation. Reprinted with permission from Ref.\ \citenum{schouten_exciton-condensate-like_2023}. Copyright 2023 The Authors, published by the American Physical Society.}
\label{fig:light_harvest}
\end{figure}

To probe the possible connection between exciton condensation and exciton transfer in the FMO complex, Schouten et\ al.\ modeled the exciton dynamics of the Fenna-Matthews-Olson (FMO) complex and probed the particle-hole RDM from the resulting density matrices for ODLRO. Electron correlation was explicitly included by using a multi-site model that allowed for the introduction of intrachromophore coupling (scaled by a coupling parameter $V$). A schematic of the equivalent interacting paths for exciton transfer created by introducing coupling is depicted in Fig.\ \ref{fig:FMO_a}.  Exciton dynamics were simulated using the Lindblad equation to account for dissipation, dephasing, and transfer to the reaction center consistent with the methodology described in Refs.\ \citenum{plenio_dephasing-assisted_2008,mazziotti_effect_2012}. The resulting dynamics (for the coupled and uncoupled models) are shown in Fig.\ \ref{fig:FMO_b}, and demonstrate the increased rate of exciton transfer for the coupled model. Calculation of $\lambda_G$ along the trajectory results in the presence of an eigenvalue greater than one, which evolves along with the exciton population transfer (Fig.\ \ref{fig:FMO_c}. (Refer to Ref.\ \citenum{schouten_exciton-condensate-like_2023} for a discussion of the method used to obtain $\lambda_G$.)

The presence of a large $\lambda_G$ that evolves with the exciton transfer dynamics of the FMO complex reveals the emergence of a coherent state of excitons connected to the energy transfer in the system. While macroscopic exciton condensation is not possible in the single-excitation manifold of this study, the presence of ODLRO in the particle-hole RDM nonetheless indicates the emergence of microscopic condensation-like behavior associated with correlation and entanglement. The results of this work demonstrate the potential involvement of an exciton-condensate-like state in enhancing exciton transfer in this system. Alongside the works observing the beginnings of exciton condensation in molecular-scale systems, this work demonstrates that the type of nonclassical long-range order associated with exciton condensation is possible outside of the macroscopic limit. Linking exciton condensation to energy transfer in light harvesting complexes supports the idea that exciton condensation should be considered as a member of a family of related phenomena involving excitonic entanglement and condensation, which can lead to highly efficient energy transfer. Another exciting implication of this work is that this understanding of the principles of highly efficient excitonic energy transfer could also be applied to design synthetic energy transport materials that use an exciton-condensate-like mechanism.

\section{Fermion-Exciton Condensates}
\begin{figure}
\centering
\hfill
\begin{subfigure}{0.55\linewidth}
\caption{}
\label{fig:fermion_exciton_a}
    \centering\includegraphics[height=0.22\textheight, keepaspectratio]{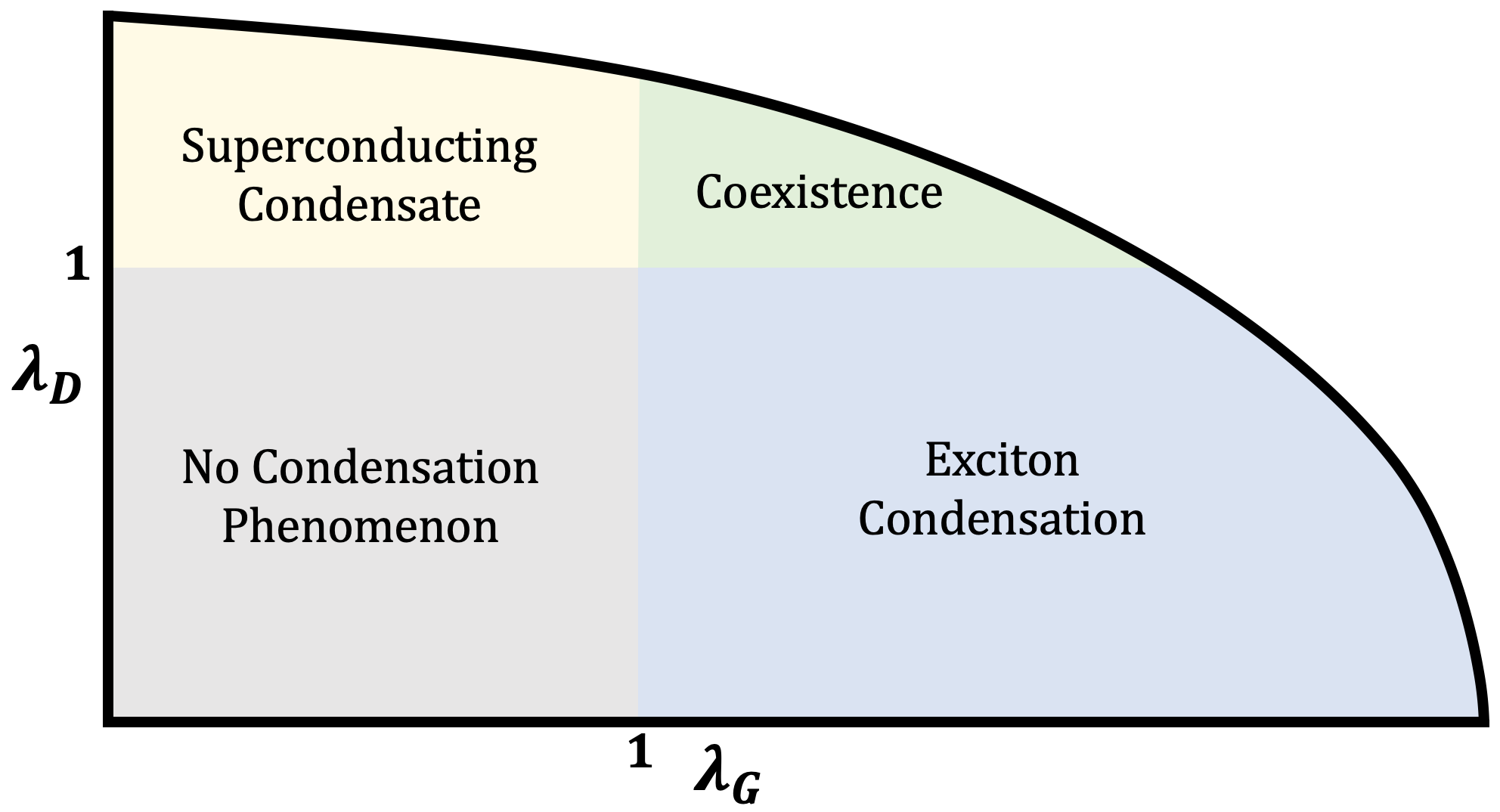}
\end{subfigure}
\hfill
\begin{subfigure}{0.43\linewidth}
\caption{}
\label{fig:fermion_exciton_b}
    \centering\includegraphics[height=0.22\textheight, keepaspectratio]{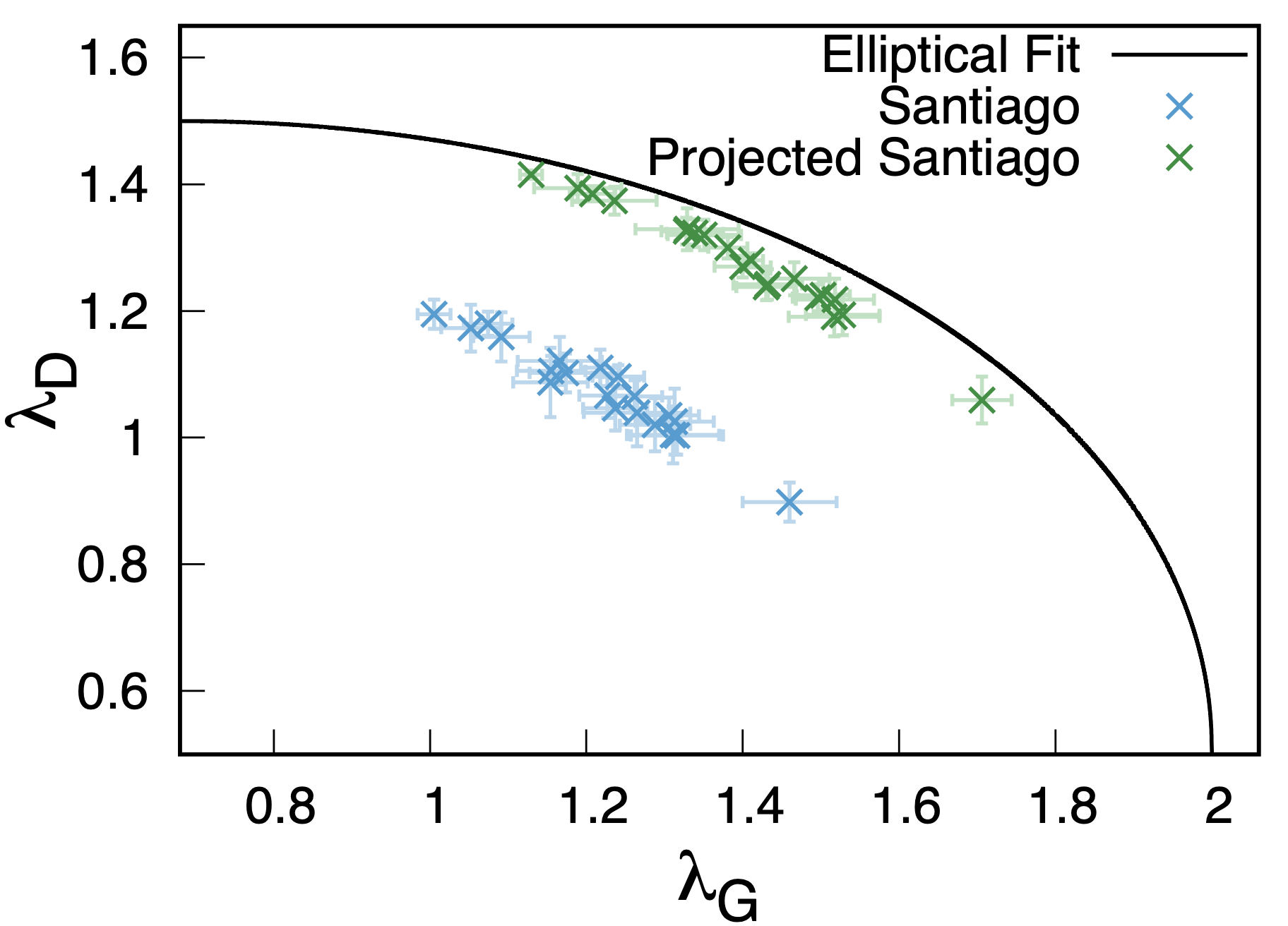}
\end{subfigure}
\hfill
\caption{(a) The condensate phase diagram in the phase space of the metrics for fermion pair condensation, $\lambda_D$, and exciton condensation, $\lambda_G$. (b) The large eigenvalues of the $^2D$ and $^2\Tilde{G}$ matrices ($\lambda_D$ and $\lambda_G$) for state preparations on IBM Quantum’s Santiago quantum computer \cite{ibm-q-team_ibm-q-5_2021} with input angles that span the region exhibiting dual condensate character plotted against the elliptical fit \cite{sager_potential_2020} obtained from the unconstrained scan of $\lambda_D$ and $\lambda_G$. The eigenvalues before error correction are plotted in blue, while the eigenvalues after error mitigation via projection of appropriate components to zero are plotted in green. Reprinted with permission from Ref.\ \citenum{sager_entangled_2022}. Copyright 2022 American Physical Society.}
\label{fig:fermion_exciton}
\end{figure}

As detailed above, bilayers of two-dimensional materials are a promising platform for the realization of exciton condensation. Additionally, superconductivity, the condensation of Cooper (electron-electron pairs), has been observed in two-dimensional layered materials.\cite{cao_unconventional_2018, yankowitz_tuning_2019, xi_ising_2016, li_nontrivial_2018, hsu_inversion-protected_2020, hao_electric_2021, zhou_isospin_2022, zhang_enhanced_2023, li_tunable_2024} The existence of both exciton condensation and fermion-pair condensation (i.e., the condensation of  fermion-fermion pairs) in such structures, as well as the mediation of superconductivity of Cooper pairs at higher temperatures via an excitonic mechanism,\cite{laussy_superconductivity_2012,skopelitis_interplay_2018} inspired Sager et\ al.\ to explore the possibility of exciton condensation and fermion-pair condensation coexisting in a single quantum state, i.e., a joint fermion-exciton condensate (FEC) state.\cite{sager_potential_2020} Such a joint FEC state may theoretically be capable of both the frictionless transfer of electrons\textemdash as in superconductivity\textemdash and the frictionless transfer of excitation energy\textemdash as in exciton condensation. As detailed in Sec.\ \hyperref[ODLRO]{2}, an eigenvalue greater than one in the particle-particle reduced density matrix, $^2D$, is characteristic of fermion-pair condensation as it indicates the presence of more than one fermion-fermion quasibosonic pair in a single quantum state (i.e., a geminal).

The potential for such a combined state was first established computationally in the limit of low particle number ($N$) and theoretically in the large-$N$ thermodynamic limit in Ref.\ \citenum{sager_potential_2020}. In each case, the existence of the combined state was established by the presence of both $\lambda_G > 1$ and $\lambda_D > 1$. The computational results in the low-$N$ limit also demonstrate an elliptical trade-off between strong exciton condensation (large $\lambda_G$) and strong fermion-pair condensation (large $\lambda_D$), as shown in Fig.\ \ref{fig:fermion_exciton_a}. The theoretical results in the large-$N$ thermodynamic limit demonstrate that fermion-exciton condensate wave functions can be constructed by entangling any fermion-pair condensate wave function ($\vert \Psi_D\rangle$) with any exciton-condensate wave function ($\vert \Psi_G\rangle$). This wavefunction is mathematically represented as
\begin{equation}
    \ket{\Psi_{FEC}} = \frac{1}{\sqrt{2 - |\Delta|}}(\ket{\Psi_D}-\text{sgn}(\Delta)\ket{\Psi_G}),
\end{equation}
\noindent where $\Delta = 2\langle\Psi_D|\Psi_G\rangle$. For details of the computational and theoretical investigations, see Ref.\  \citenum{sager_potential_2020}. The knowledge that fermion-exciton condensate wavefunctions can be constructed by entangling fermion pair and exciton condensate wavefunctions was later used to explore fermion-exciton condensation with a model Hamiltonian.\cite{sager_simultaneous_2022} The resulting wavefunction is an entanglement of BCS-like superconducting wavefunctions and Lipkin-like exciton condensate wavefunctions, which is consistent with the prediction from Ref.\  \citenum{sager_potential_2020}.

After theoretically and computationally establishing the potential for a fermion-exciton condensate, Sager et al. demonstrated the physical realization of fermion-exciton condensation on a quantum computer.\cite{sager_entangled_2022} As quantum devices had recently been used for the study of strongly correlated quantum matter,\cite{ma_dissipatively_2019, sager_preparation_2020} quantum computing provided a promising platform. Quantum computers provide an ideal venue for studying strongly correlated quantum systems such as exciton condensates, as qubits allow for the preparation of a quantum state with precisely controllable interactions and dynamics.\cite{ma_dissipatively_2019, sager_preparation_2020} Quantum devices had been utilized previously by Sager-Smith et. al. to prepare both exciton condensates and condensates of Cooper pairs via superconducting transmon qubits.\cite{sager_preparation_2020, sager_cooper-pair_2022} Thus, using quantum state preparations utilized in these prior studies, one can construct a state preparation for a simultaneous fermion-exciton condensate on a quantum computer.\cite{sager_simultaneous_2022} Applying a state preparation that represents the entanglement of the fermion-pair condensate and exciton condensate quantum states and scanning over input angle parameters to span the space of entangled states, results in quantum states with both fermion-pair condensate and exciton condensate character (as evidenced by simultaneous $\lambda_D>1$ and $\lambda_G>1$). These results are shown in Fig.\ \ref{fig:fermion_exciton_b} for states prepared on IBM Quantum's Santiago \cite{ibm-q-team_ibm-q-5_2021} quantum device. Current quantum devices are subject to high levels of error-causing noise, which results in the deviation of the eigenvalues from the theoretical ideal values. However, applying error correction (by projecting to zero the  contributions to the density matrices from the qubit basis states that are not expected to contribute) results in eigenvalues that closely approximate the ideal $\lambda_D$ and $\lambda_G$s. These results suggest that the realm of condensation phenomena may be richer than previously thought, with exciton condensation being possible in states beyond traditional or pure exciton condensates. In the three papers, FECs were theoretically predicted, computationally modeled, and demonstrated on a physical quantum device, motivating future exploration of such joint condensate states in physical systems. As both fermion-pair condensates and exciton condensates have been found in bilayer materials, such materials are promising candidates for the realization of a fermion-exciton condensate state.

\section{Conclusions and Outlook}
In this perspective, we highlight how the eigenvalue metric for exciton condensation, $\lambda_G$, has enabled new advances in studying exciton condensation from a chemical perspective. We demonstrate how viewing exciton condensation through the lens of off-diagonal long range order can help bridge connections between disparate perspectives on exciton condensation, as well as unite the study of exciton condensates to the broader field of exciton phenomena involving strong correlation and highly efficient energy transport. $\lambda_G$ is a particularly valuable metric because it reflects the strong correlation that underlies condensation phenomena, and can ascertain the presence of ``exciton-condensate-like'' coherent states of excitons even outside of the ideal condensate limit. These coherent states include molecular-scale exciton condensates in stacked molecules and amorphous materials as well as exciton-condensate-like mechanisms in highly efficient exciton transfer.

The existence of various emergent condensate-like states of collective excitons suggests that macroscopic exciton condensation can be understood as an extreme case in the general family of exciton phenomena involving strong correlation and highly efficient energy transport. These exciton-condensate-like phenomena may be possible under conditions more easily realizable than those necessary for macroscopic exciton condensation, while still facilitating highly efficient energy transfer through strong correlation and entanglement. The possibility of a joint fermion-exciton condensate state is also a very exciting avenue that motivates future investigations, as such a state would theoretically be capable of both the frictionless transfer of electrons and of excitation energy. Furthermore, while not discussed in detail here except in the case of dual fermion-exciton condensation, $\lambda_G$ is readily extracted for states prepared on quantum devices. This makes quantum computers a promising platform for the study of exciton condensation, and studies are ongoing involving the preparation and characterization of exciton condensate states with superconducting transmon qubits. Finally, recent work has employed the cumulant eigenvalue metric for exciton condensation as an entanglement witness, with the elements of the cumulant obtained from solid-state spectroscopy, suggesting potential applications for the cumulant eigenvalue metric for exciton condensation in experimental contexts.

Together, these promising results suggest that the eigenvalue metric for exciton condensation will continue to be a vital tool in discovering avenues towards realizing ultra-efficient energy transport through strongly correlated excitons at ambient conditions, both in the exciton condensate limit and beyond.
\begin{acknowledgement}

D.A.M. gratefully acknowledges support from the U. S. National Science Foundation Grant Nos. CHE-2155082 and the Department of Energy, Office of Basic Energy Sciences, Grant DE-SC0019215.  We acknowledge the use of IBM Quantum services for this work. The views expressed are those of the authors, and do not reflect the official policy or position of IBM or the IBM Quantum team.
\end{acknowledgement}

	\bibliography{references}


\end{document}